\documentclass{sig-alternate}

\makeatletter
\newif\if@restonecol
\makeatother

\usepackage[linesnumbered, ruled]{algorithm2e}
\usepackage{graphicx}
\usepackage{epstopdf}
\usepackage{subfig}
\usepackage{url}

\begin{document}

\title{High Performance Risk Aggregation:}
\subtitle{Addressing the Data Processing Challenge the Hadoop MapReduce Way
}

\author{
\alignauthor
Z. Yao, B. Varghese\titlenote{Corresponding author. E-mail: varghese@cs.dal.ca. Webpage: \url{http://www.blessonv.com}}and A. Rau-Chaplin\\
	\affaddr{Faculty of Computer Science, Dalhousie University, Halifax, Canada}\\
	\email{\{yao, varghese, arc\}@cs.dal.ca}
}

\date{10 April 2013}

\maketitle
\begin{abstract}
Monte Carlo simulations employed for the analysis of portfolios of catastrophic risk process large volumes of data. Often times these simulations are not performed in real-time scenarios as they are slow and consume large data. Such simulations can benefit from a framework that exploits parallelism for addressing the computational challenge and facilitates a distributed file system for addressing the data challenge. To this end, the Apache Hadoop framework is chosen for the simulation reported in this paper so that the computational challenge can be tackled using the MapReduce model and the data challenge can be addressed using the Hadoop Distributed File System. A parallel algorithm for the analysis of aggregate risk is proposed and implemented using the MapReduce model in this paper. An evaluation of the performance of the algorithm indicates that the Hadoop MapReduce model offers a framework for processing large data in aggregate risk analysis. A simulation of aggregate risk employing 100,000 trials with 1000 catastrophic events per trial on a typical exposure set and contract structure is performed on multiple worker nodes in less than 6 minutes. The result indicates the scope and feasibility of MapReduce for tackling the computational and data challenge in the analysis of aggregate risk for real-time use. 
\end{abstract}




\section{Introduction}

In the domain of large-scale computational analysis of risk, large amounts of data need to be rapidly processed and millions of simulations need to be quickly performed (for example, \cite{1, 2, 2a}). This can be achieved only if data is efficiently managed and parallelism is exploited within algorithms employed in the simulations. The domain, therefore, inherently opens avenues to exploit the synergy that can be achieved by bringing together state-of-the-art techniques in data processing and management and high-performance computing. Research on aggregate analysis of risks \cite{s1, s2, s4} using high-performance computing is sparse at best. The research reported in this paper is motivated towards exploring techniques for employing high-performance computing not only to speed up the simulation but also to process and manage data efficiently for the aggregate analysis of risk. In this context the MapReduce model \cite{3, 4, 5} is used for achieving high-performance aggregate analysis of risks. 

The aggregate analysis of risk is a Monte Carlo simulation performed on a portfolio of risks that an insurer or reinsurer holds. A portfolio can cover risks related to catastrophic events such as earthquakes, floods or hurricanes, and may comprise tens of thousands of contracts. The contracts generally follow an `eXcess of Loss' (XL) \cite{6, 7} structure providing coverage for single event occurrences or multiple event occurrences, or a combination of both single and multiple event occurrences. Each trial in the aggregate analysis simulation represents a view of the occurrence of catastrophic events and the order in which they occur within a contractual year. The trial also provides information on how the occurrence of an event in a contractual year will interact with complex treaty terms to produce an aggregated loss. A pre-simulated Year Event Table (YET) containing between several thousands and millions of alternative views of a single contractual year is input for the aggregate analysis. The output of aggregate analysis is a Year Loss Table (YLT). From a YLT, an insurer or a reinsurer can derive important portfolio risk metrics such as the Probable Maximum Loss (PML) \cite{8, 9} and the Tail Value-at-Risk (TVaR) \cite{10, 11} which are used for both internal risk management and reporting to regulators and rating agencies. 

In this paper, the analysis of portfolios of catastrophic risk is proposed and implemented using a MapReduce model on the Hadoop \cite{12, 13, 14} platform. The algorithm rapidly consumes large amounts of data in the form of the YET and Event Loss Tables (ELT). Therefore, the challenges of organising input data and processing it efficiently, and applying parallelism within the algorithm are considered. The MapReduce model lends itself well towards solving embarrassingly parallel problems such as the aggregate analysis of risk, and is hence chosen to implement the algorithm. The algorithm employs two MapReduce rounds to perform both the numerical computations as well as to manage and process data efficiently. The algorithm is implemented on the Apache Hadoop platform. The Hadoop Distributed File System (HDFS) and the Distributed Cache (DC) are key components offered by the Hadoop platform in addressing the data challenges. The preliminary results obtained from the experiments of the analysis indicate that the MapReduce model can be used to scale the analysis over multiple nodes of a cluster; parallelism can be exploited in the analysis for achieving faster numerical computations and data management. 

The remainder of this paper is organised as follows. Section \ref{ara} considers the sequential and MapReduce algorithm for the analysis of aggregate risk. Section \ref{hadoop} presents the implementation of the MapReduce algorithm on the Apache Hadoop Platform and the preliminary results obtained from the experimental studies. Section \ref{conclusion} concludes this paper by considering future work. 

\section{Analysis of Aggregate Risk}
\label{ara}

The sequential and MapReduce algorithm of the analysis of aggregate risk is presented in this section. There are three inputs to the algorithm for the analysis of aggregate risk, namely the $YET$, $PF$, and a pool of $ELTs$. The $YET$ is the Year Event Table which is the representation of a pre-simulated occurrence of Events $E$ in the form of trials $T$. Each Trial captures the sequence of the occurrences of Events for a year using Time-stamps in the form of Event Time-stamp pairs. The $PF$ is a portfolio that represents a group of Programs, $P$, which in turn represents a set of Layers, $L$ that covers a set of $ELTs$ using financial terms. The $ELT$ is the Event Loss Table which represents the losses that correspond to an event based on an exposure (one event can appear over different ELTs with different losses). 

The intermediary output of the algorithm are the Layer Loss Table $LLT$ consisting Trial-Loss pairs. The final output of the algorithm is $YLT$, which is the Year Loss Table that contains the losses covered by a portfolio.

\subsection{Sequential Algorithm}
\label{sequential}

Algorithm \ref{algorithm1} shows the sequential analysis of aggregate risk. The algorithm scans through the hierarchy of the portfolio, $PF$; firstly through the Programs, $P$, followed by the Layers, $L$, then the Event Loss Tables, $ELTs$. Line nos. 5-8 shows how the loss associated with an Event in the $ELT$ is computed. For this, the loss, $l_{E}$ associated with an Event, $E$ is retrieved, after which contractual financial terms to the benefit of the Layer are applied to the losses and are summed up as $l'_{E}$.

\begin{algorithm} 
\caption{Sequential Aggregate Risk Analysis}
\label{algorithm1}
\SetAlgoLined
\DontPrintSemicolon

\SetKwInOut{Input}{Input}
\SetKwInOut{Output}{Output}

\Input{$YET$, $ELT$ pool, $PF$}
\Output{$YLT$}

\BlankLine

\For{each Program, $P$}{
	\For{each Layer, $L$, in $P$}{
		\For{each Trial, $T$, in $YET$}{
			\For{each Event, $E$, in $T$}{
				\For{each $ELT$ covered by $L$}{
					Lookup $E$ in the $ELT$ and find corresponding loss, $l_{E}$\;
					$l'_{E} \leftarrow$ $l'_{E}$ + $l_{E}$\;
				}
				Apply Occurrence Financial Terms to $l'_{E}$\;	
				$l_{T} \leftarrow$ $l_{T}$ + $l'_{E}$\;
			}
			Apply Aggregate Financial Terms to $l_{T}$\;
			Populate Trial-Loss pairs in $LLT$ using $l_{T}$\; 	
		}
	}
	Sum losses of Trial-Loss pairs in all $LLT$\;
	Populate Trial-Loss pairs in $PLT$\;	
}			
Aggregate losses of Trial-Loss pairs in $PLT$\;
Populate $YLT$\;
\end{algorithm}

In line nos. 9 and 10, two Occurrence Financial Terms, namely the Occurrence Retention and the Occurrence Limit are applied to the loss, $l'_{E}$ and summed up as $l_{T}$. The $l_{T}$ losses correspond to the total loss in one trial. Occurrence Retention refers to the retention or deductible of the insured for an individual occurrence loss, where as Occurrence Limit refers to the limit or coverage the insurer will pay for occurrence losses in excess of the retention. The Occurrence Financial Terms capture specific contractual properties of 'eXcess of Loss' treaties as they apply to individual event occurrences only.

In line nos. 12 and 13, two Aggregate Financial Terms, namely the Aggregate Retention and the Aggregate Limit are applied to the loss, $l_{T}$ to produce aggregated loss for a Trial. Aggregate Retention refers to the retention or deductible of the insured for an annual cumulative loss, where as Aggregate Limit refers to the limit or coverage the insurer will pay for annual cumulative losses in excess of the aggregate retention. The Aggregate Financial terms captures contractual properties as they apply to multiple event occurrences. The Trial-Loss pairs are then used to populate Layer Loss Tables $LLTs$; each Layer is represented using a Layer Loss Table consisting of Trial-Loss pairs.

In line nos. 16 and 17, the trial losses are aggregated from the Layer level to the Program level. The losses are represented again as a Trial-Loss pair and are used to populate Program Loss Tables $PLTs$; each Program is represented using a Program Loss Table.

In line nos. 19 and 20, the trial losses are aggregated from the Program level to the Portfolio level. The Trial-Loss pairs are populated in the Year Loss Table $YLT$ which represents the output of the analysis of aggregate risk. Financial functions or filters are then applied on the aggregate loss values.

\subsection{MapReduce Algorithm}
\label{mapreduce}

\begin{algorithm} 
\caption{Parallel Aggregate Risk Analysis}
\label{algorithm2}
\SetAlgoLined
\DontPrintSemicolon

\SetKwInOut{Input}{Input}
\SetKwInOut{Output}{Output}

\Input{$YET$, $ELT$ pool, $PF$}
\Output{$YLT$}

\BlankLine
\ForAll{Programs of $P$}{
	\ForAll{Layers $L$ in P}{
			$LLT \leftarrow MapReduce_{1}$($L$, $YET$)\;
	}
}
$YLT \leftarrow MapReduce_{2}$($LLTs$)\;
\end{algorithm}

MapReduce is a programming model developed by Google for processing large amount of data on large clusters. A map and a reduce function are adopted in this model to execute a problem that can be decomposed into sub-problems with no dependencies; therefore the model is most attractive for embarrassingly parallel problems. This model is scalable across large number of computing resources. In addition to the computations, the fault tolerance of the execution, for example, handling machine failures are taken care by the MapReduce model. An open-source software framework that supports the MapReduce model, Apache Hadoop \cite{12, 13, 14}, is used in the research reported in this paper.

The MapReduce model lends itself well towards solving embarrassingly parallel problems, and therefore, the analysis of aggregate risk is explored on MapReduce. In the analysis of aggregate risks, the Programs contained in the Portfolio are independent of each other, the Layers contained in a Program are independent of each other and further the Trials in the Year Event Table are independent of each other. This indicates that the problem of analysing aggregate risks requires a large number of computations which can be performed as independent parallel problems.

Another reason of choice for the MapReduce model is that it can handle large data processing for the analysis of aggregate risks. For example, consider a Year Event Table comprising one million simulations, which is approximately 30 GB. So for a Portfolio comprising 2 Programs, each with 10 Layers, the approximate volume of data that needs to be processed is 600 GB.

Further MapReduce implementations such as Hadoop provide dynamic job scheduling based on the availability of cluster resources and distributed file system fault tolerance.

Algorithm \ref{algorithm2} shows the MapReduce analysis of aggregate risk. The aim of this algorithm is similar to the sequential algorithm in which the algorithm scans through the Portfolio, $PF$; firstly through the Programs, $P$, and then through the Layers, $L$. The first round of MapReduce jobs, denoted as $MapReduce_{1}$ are launched for all the Layers. The Map function (refer Algorithm \ref{algorithm3}) scans through all the Event Loss Tables $ELTs$ covered by the Layers $L$ to compute the losses $l'_{E}$ in parallel for every Event in the ELT. The computations of loss $l_{T}$ at the Layer level are performed in parallel by the Reduce function (refer Algorithm \ref{algorithm4}). The output of $MapReduce_{1}$ is a Layer Loss Table $LLT$.

The second round of MapReduce jobs, denoted as \\$MapReduce_{2}$ are launched for aggregating all the $LLTs$ in each Program to a $YLT$. 

\begin{algorithm} 
\caption{Map function in $MapReduce_{1}$ of the Analysis of Aggregate Risk}
\label{algorithm3}
\SetAlgoLined
\DontPrintSemicolon

\SetKwInOut{Input}{Input}
\SetKwInOut{Output}{Output}

\Input{$<T$, $E>$}
\Output{$<T$, $l'_{E}>$}

\BlankLine
		\For{each $ELT$ covered by $L$}{
			Lookup $E$ in the $ELT$ and find corresponding loss, $l_{E}$\;
			Apply Financial Terms to $l_{E}$\;
			$l'_{E} \leftarrow$ $l'_{E}$ + $l_{E}$\;
		}
		Emit($T$, $l'_{E}$)
\end{algorithm}

\begin{algorithm} 
\caption{Reduce Function in $MapReduce_{1}$ of the Analysis of Aggregate Risk}
\label{algorithm4}
\SetAlgoLined
\DontPrintSemicolon

\SetKwInOut{Input}{Input}
\SetKwInOut{Output}{Output}

\Input{$T$, $L'_{E}$}
\Output{$<T$, $l_{T}>$}

\BlankLine
	\For{each $l'_{E}$ in $L'_{E}$}{
		Apply Occurrence Financial Terms to $l'_{E}$\;	
		$l_{T} \leftarrow$ $l_{T}$ + $l'_{E}$\;	
	}
	Apply Aggregate Financial Terms to $l_{T}$\;
	Emit($T$, $l_{T}$)

\end{algorithm}

The master node of the cluster solving a problem partitions the input data to intermediate files effectively splitting the problem into sub-problems. The sub-problems are distributed to the worker nodes by the master node, often referred to as the `Map' step performed by the Mapper. The map function executed by the Mapper receives as input a $<key, value>$ pair to generate a set of $<intermediate$ $key, intermediate$ $value>$ pairs. The results of the decomposed sub-problems are then combined by the Reducer referred to as the `Reduce' step. The Reduce function executed by each Reducer merges the $<intermediate$ $key, intermediate$ $value>$ pairs to generate a final output. The Reduce function receives all the values corresponding to the same intermediate key.

Algorithm \ref{algorithm3} and Algorithm \ref{algorithm4} show how parallelism is achieved by using the Map and Reduce functions in a first round at the Layer level. Algorithm \ref{algorithm3} shows the Map function whose inputs are a set of $T, E$ from the $YET$, and the output is a Trial-Loss pair $<T,l'_{E}>$ which corresponds to an Event. To estimate the loss, it is necessary to scan through every Event Loss Table $ELT$ covered by a Layer $L$ (line nos. 1-5). The loss, $l_{E}$ associated with an Event, $E$ in the $ELT$ is fetched from memory in line no. 2. Contractual financial terms to the benefit of the Layer are applied to the losses (line no. 3) to aggregate the losses as $l'_{E}$ (line no. 4). The loss for every Event in a Trial is emitted as $<T,l'_{E}>$.

Algorithm \ref{algorithm4} shows the Reduce function in the first MapReduce round. The inputs are the Trial $T$ and the set of losses ($l'_{E}$) corresponding to that Trial, represented as $L'_{E}$, and the output is a Trial-Loss pair $<T,l_{T}>$. For every loss value $l'_{E}$ in the set of losses $L'_{E}$, the Occurence Financial Terms, namely Occurrence Retention and the Occurrence Limit, are applied to $l'_{E}$ (line no. 2) and summed up as $l_{T}$ (line no. 3). The Aggregate Financial Terms, namely Aggregate Retention and Aggregate Limit are applied to $l_{T}$ (line no. 5). The aggregated loss for a Trial, $l_{T}$ is emitted as $<T, l_{T}>$ to populate the Layer Loss Table.

Algorithm \ref{algorithm5} and Algorithm \ref{algorithm6} show how parallelism is achieved by using the Map and Reduce functions in a second round for aggregating all Layer Loss Tables to produce the $YLT$. Algorithm \ref{algorithm5} shows the Map function whose inputs are a set of Layer Loss Tables $LLTs$, and the output is a Trial-Loss pair $<T,l_{T}>$ which corresponds to the Layer-wise loss for Trial $T$.

\begin{algorithm} 
\caption{Map function in $MapReduce_{2}$ of the Analysis of Aggregate Risk}
\label{algorithm5}
\SetAlgoLined
\DontPrintSemicolon

\SetKwInOut{Input}{Input}
\SetKwInOut{Output}{Output}

\Input{$LLTs$}
\Output{$<T$, $l_{T}>$}

\BlankLine
	\For{each $T$ in $LLT$}{
		Emit($<T, l_{T}>$)
	}
\end{algorithm}

\begin{algorithm} 
\caption{Reduce function in $MapReduce_{2}$ of the Analysis of Aggregate Risk}
\label{algorithm6}
\SetAlgoLined
\DontPrintSemicolon

\SetKwInOut{Input}{Input}
\SetKwInOut{Output}{Output}

\Input{$<T, L_{T}>$}
\Output{$<T$, $l'_{T}>$}

\BlankLine
	\For{each $l_{T}$ in $L_{T}$}{
		$l'_{T} \leftarrow l'_{T} + l_{T}$
	}
	Emit($<T, l'_{T}>$)
\end{algorithm}

Algorithm \ref{algorithm6} shows the Reduce function whose inputs are a set of losses corresponding to a Trial in all Layers $L_{T}$, and the output is a Trial-Loss pair $<T,l'_{T}>$ which is an entry to populate the final output, the Year Loss Table $YLT$. The function sums up trial losses $l_{T}$ across all Layers to produce a portfolio-wise aggregate loss $l'_{T}$.

\section{Implementation and Experiments on the Hadoop Platform}
\label{hadoop}

The experimental platform for implementing the MapReduce algorithm is a heterogeneous cluster comprising (a) a master node which is an IBM blade of two XEON 2.67 GHz processors comprising six cores, memory of 20 GB per processor and a hard drive of 500 GB with an additional 7 TB RAID array, and (b) six worker nodes each with an Opteron Dual Core 2216 2.4 GHz processor comprising four cores, memory of 4 GB RAM and a hard drive of 150 GB. The nodes are interconnected via Infiniband.

Apache Hadoop, an open-source software framework is used for implementing the MapReduce analysis of aggregate risk. Other available frameworks \cite{15, 16} require the use of additional interfaces, commercial or web-based, for deploying an application and were therefore not chosen.

The Hadoop framework works in the following way for a MapReduce round. First of all the data files from the Hadoop Distributed File System (HDFS) is loaded using the \texttt{InputFormat} interface. HDFS provides a functionality called Distributed Cache for distributing small data files which are shared by the nodes of the cluster. The Distributed Cache provides local access to shared data. The \texttt{InputFormat} interface specifies the input the Mapper and splits the input data as required by the Mapper. The \texttt{Mapper} interface receives the partitioned data and emits intermediate key-value pairs. The \texttt{Partitioner} interface receives the intermediate key-value pairs and controls the partitioning of these keys for the \texttt{Reducer} interface. Then the \texttt{Reducer} interface receives the partitioned intermediate key-value pairs and generates the final output of this MapReduce round. The output is received by the \texttt{OutputFormat} interface and provides it back to HDFS.

\begin{figure*} 
\centering
	\subfloat[First MapReduce round]{\label{fig:s11}\includegraphics[width=0.5\textwidth]{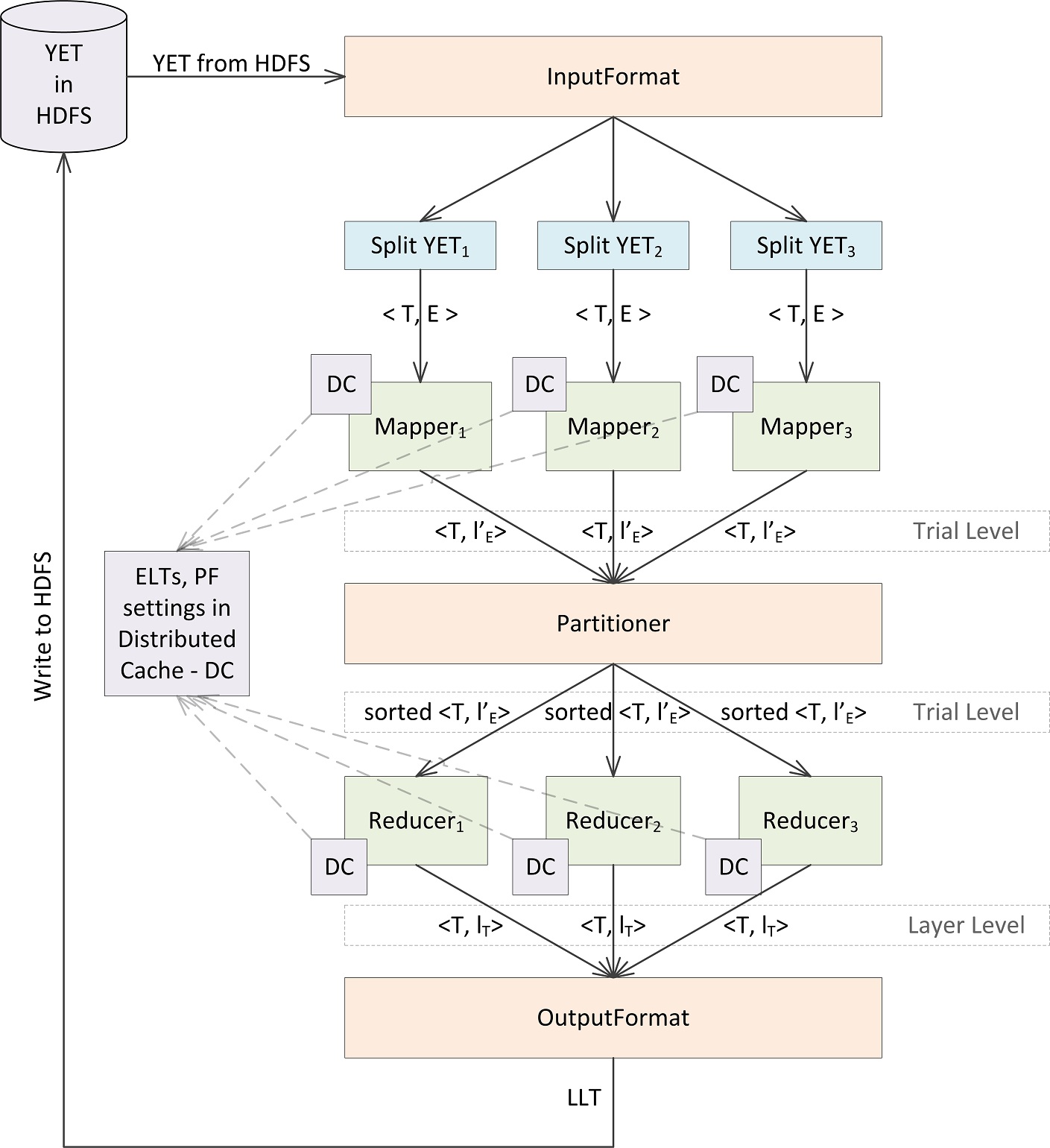}} \hfill
	\subfloat[Second MapReduce round]{\label{fig:s12}\includegraphics[width=0.5\textwidth]{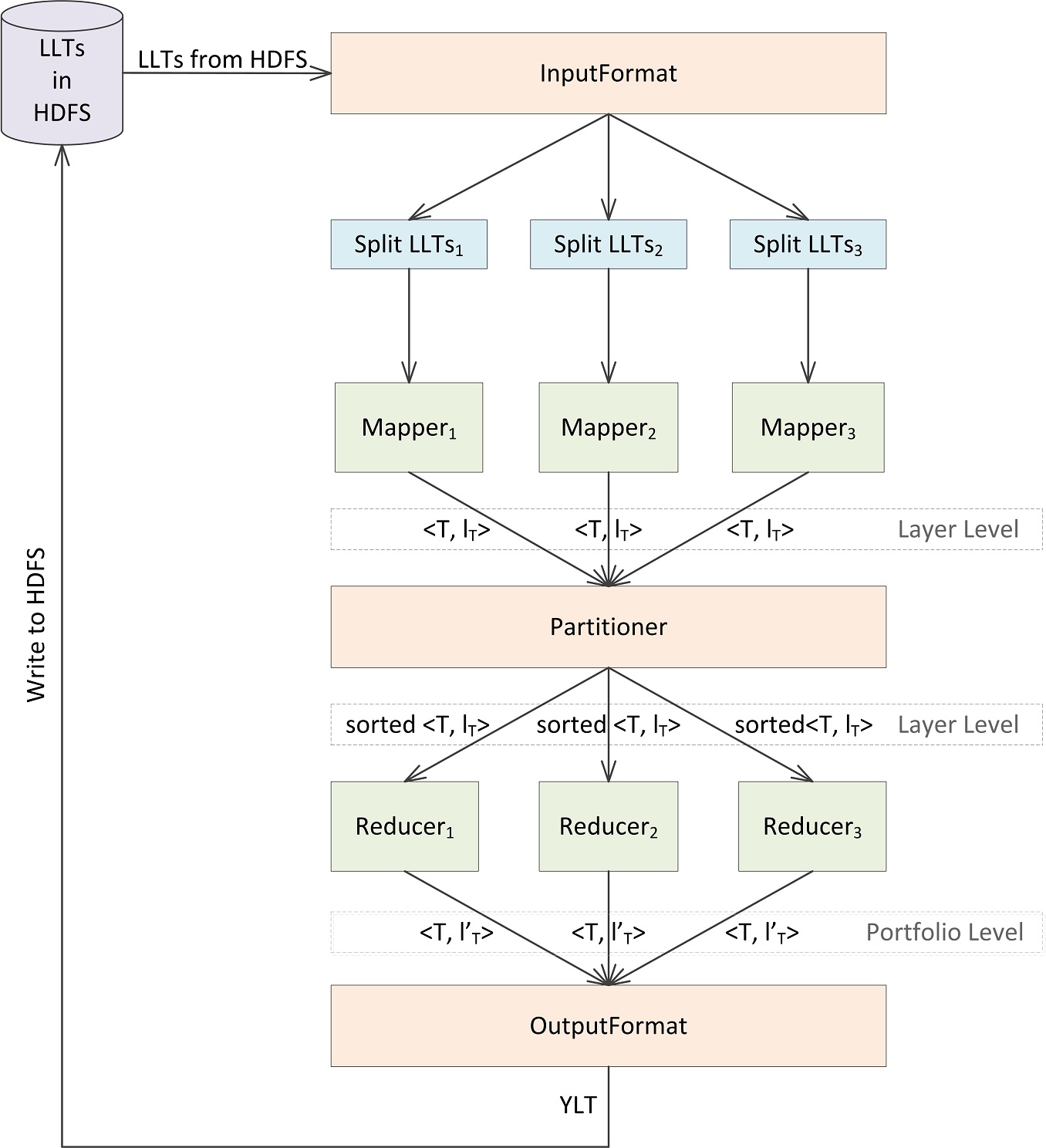}} \\
\caption{MapReduce rounds in the Hadoop implementation}
\label{figure1}
\end{figure*}

The input data for MapReduce ARA which are the Year Event Table $YET$, the pool of Event Loss Table $ELT$ and the Portfolio $PF$ specification are stored on HDFS. The master node executes Algorithm \ref{algorithm2} to generate the Year Loss Table $YLT$ which is again stored on the HDFS. The two MapReduce rounds are illustrated in Figure \ref{figure1}.

In the first MapReduce round the \texttt{InputFormat} interface splits the $YET$ based on the number of Mappers specified for the MapReduce round. The Mappers are configured such that they also receive the $ELTs$ covered by one Layer which are contained in the distributed cache. The Mapper applies Financial Terms to the losses. In this implementation combining the $ELTs$ is considered for achieving fast lookup. A typical $ELT$ would contain entries in the form of an Event ID and related loss information. When the $ELTs$ are combined they contain an Event ID and the loss information related to all the individual $ELTs$. This reduces the number of lookups for retrieving loss information related to an Event when the Events in a Trial contained in the $YET$ are scanned through by the Mapper. The Mapper emits a trial-Event Loss pair which is collected by the Partitioner. The Partitioner delivers the Trial-Event Loss pairs to the Reducers; one Reducer receives all the Trial-Event Loss pairs related to a specific trial. The Reducer applies the Occurrence Financial and Aggregate Financial Terms to the losses emitted to it by the Mapper. Then the \texttt{OutputFormat} writes the output of the first MapReduce round as Layer Loss Tables $LLT$ to the HDFS.

In the second MapReduce round the \texttt{InputFormat} receives all the $LLTs$ from HDFS. The \texttt{InputFormat} interface splits the set of $LLTs$ and distributes them to the Mappers. The \texttt{Mapper} interface emits Layer-wise Trial-Loss pairs. The \texttt{Partitioner} receives all the Trial-Loss pairs and partitions them based on the Trial for each Reducer. The \texttt{Reducer} interface uses the partitioned Trial-Loss pairs and combines them to Portfolio-wise Trial-Loss pairs. Then the \texttt{OutputFormat} writes the output of the second MapReduce round as a Year Loss Table $YLT$ to the HDFS.

\subsection{Results}

Experiments were performed for one Portfolio comprising one Program and one Layer and sixteen Event Loss Tables. The Year Event Table has 100,000 Trials, with each Trial comprising 1000 Events. The experiments are performed for up to 12 workers as there are 12 cores available on the cluster employed for the experiments.

\begin{figure} [t] 
	\subfloat[First MapReduce round]{\label{fig:s21}\includegraphics[width=0.495\textwidth]{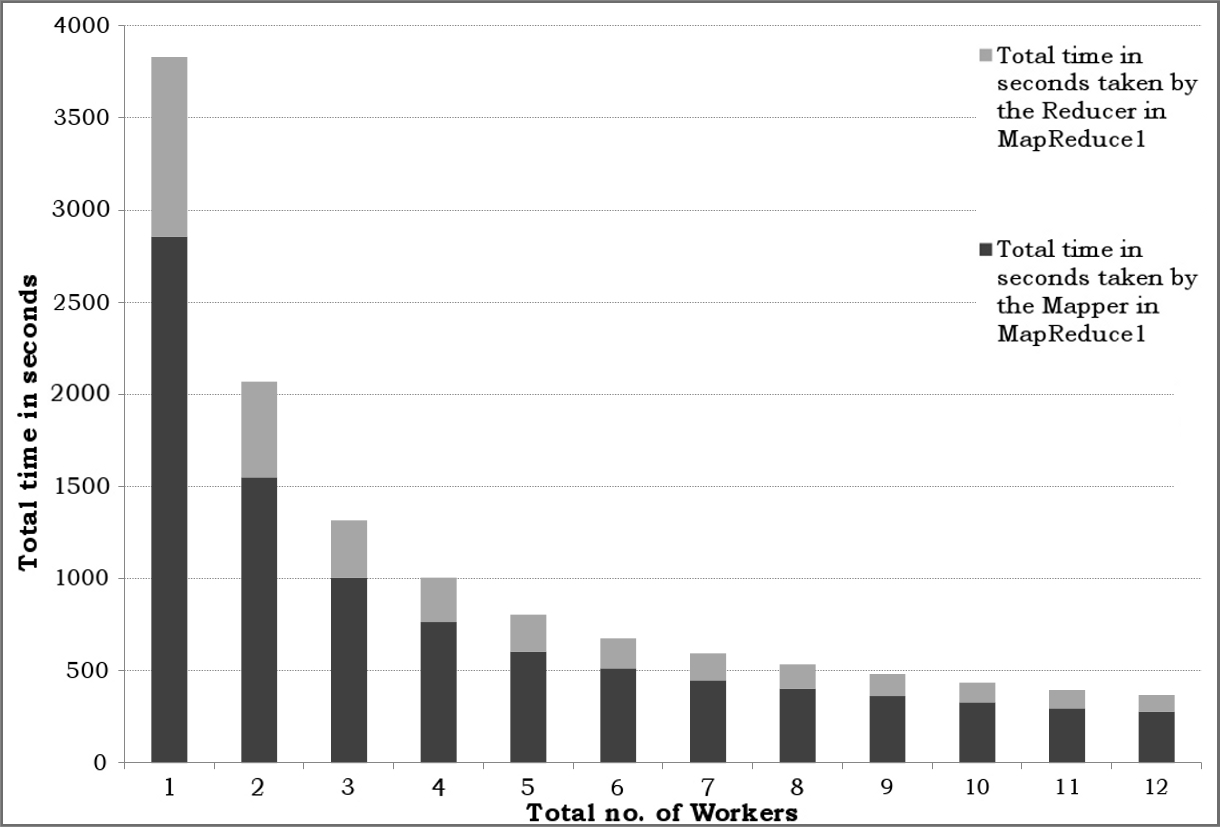}} \\
	\subfloat[Second MapReduce round]{\label{fig:s22}\includegraphics[width=0.495\textwidth]{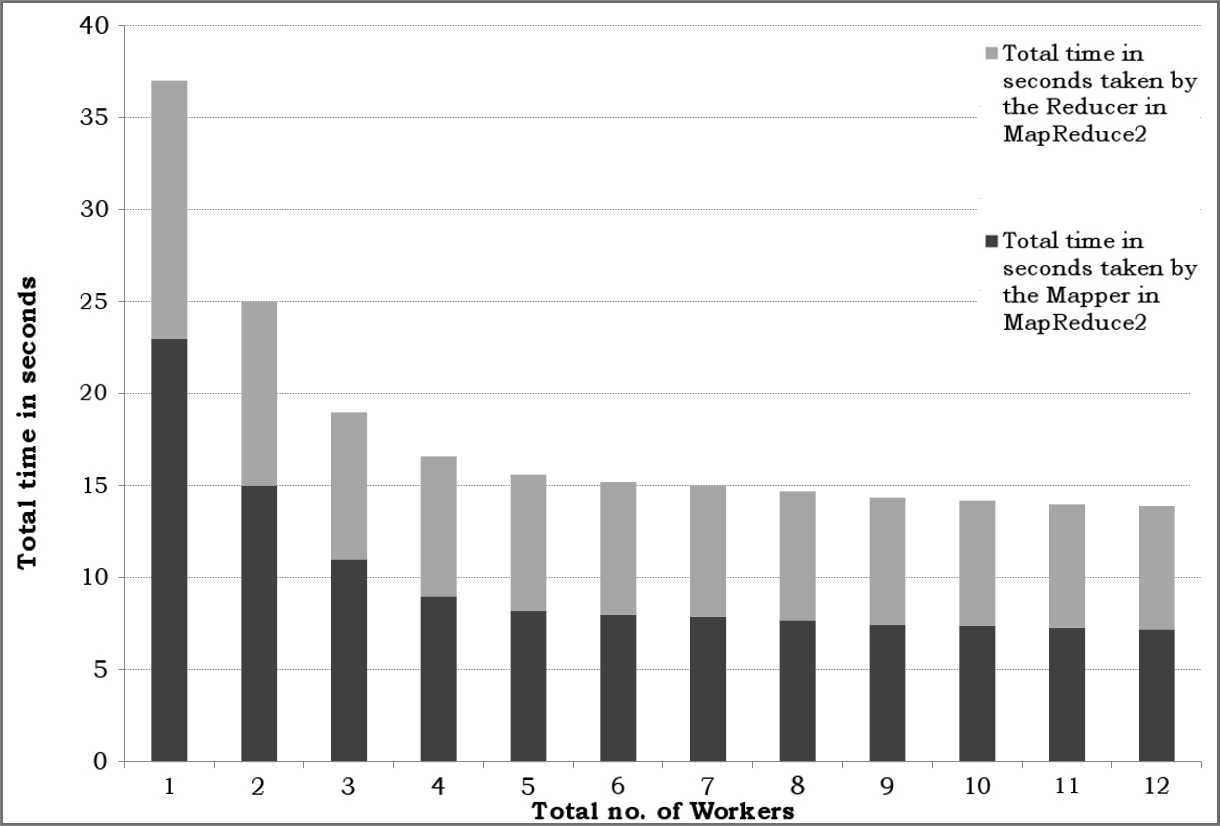}} \\
\caption{Number of workers vs total time taken in seconds for the MapReduce rounds in the Hadoop implementation}
\label{figure2}
\end{figure}

Figure \ref{figure2} shows two bar graphs for the total time taken in seconds for the MapReduce rounds when the workers are varied between 1 and 12; Figure \ref{fig:s21} for the first MapReduce round and Figure \ref{fig:s22} for the second MapReduce round. In the first MapReduce round the best timing performance is achieved on 12 Mappers and 12 Reducers taking a total of 370 seconds, with 280 seconds for the Mapper and 90 seconds for the Reducer. Over 85\% efficiency is achieved in each case using multiple worker nodes compared to 1 worker. This round is most efficient on 3 workers achieving an efficiency of 97\% and the performance deteriorates beyond the use of four workers on the cluster employed. In the second MapReduce round the best timing performance is achieved again on 12 Mapper and 12 Reducers taking a total of 13.9 seconds, with 7.2 seconds for the Mapper and 6.7 seconds for the Reducer. Using 2 workers has the best efficiency of 74\%; the efficiency deteriorates beyond this. The second MapReduce round has performed poorly compared to the first round as there are large I/O and initialisation overheads on the workers.

\begin{figure} [t] 
\centering
	\subfloat[No. of Mappers vs Time time taken for (a) applying Financial Terms, (b) local I/O operation by each Mapper, and (c) data delivery]{\label{fig:s31}\includegraphics[width=0.495\textwidth]{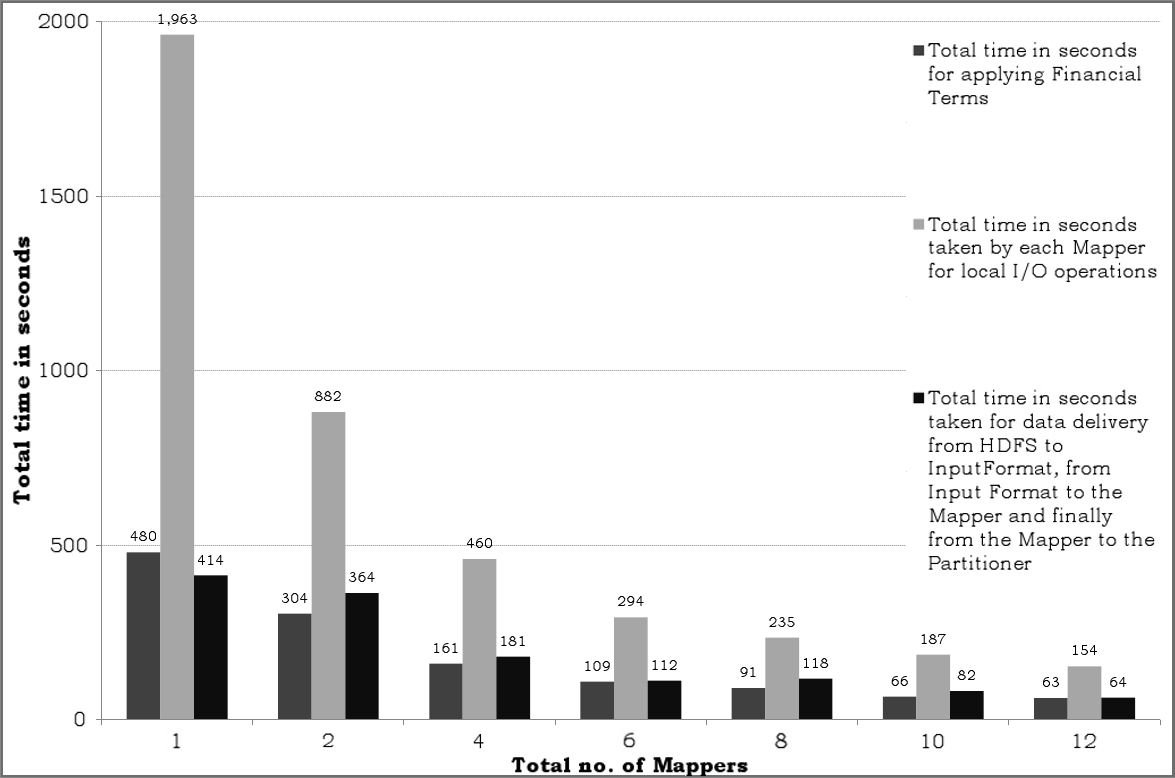}} \hfill
	\subfloat[No. of Reducers vs Time time taken for (a) applying Occurrence and Aggregate Financial Terms, (b) local I/O operation by each Reducer, and (c) data delivery]{\label{fig:s32}\includegraphics[width=0.495\textwidth]{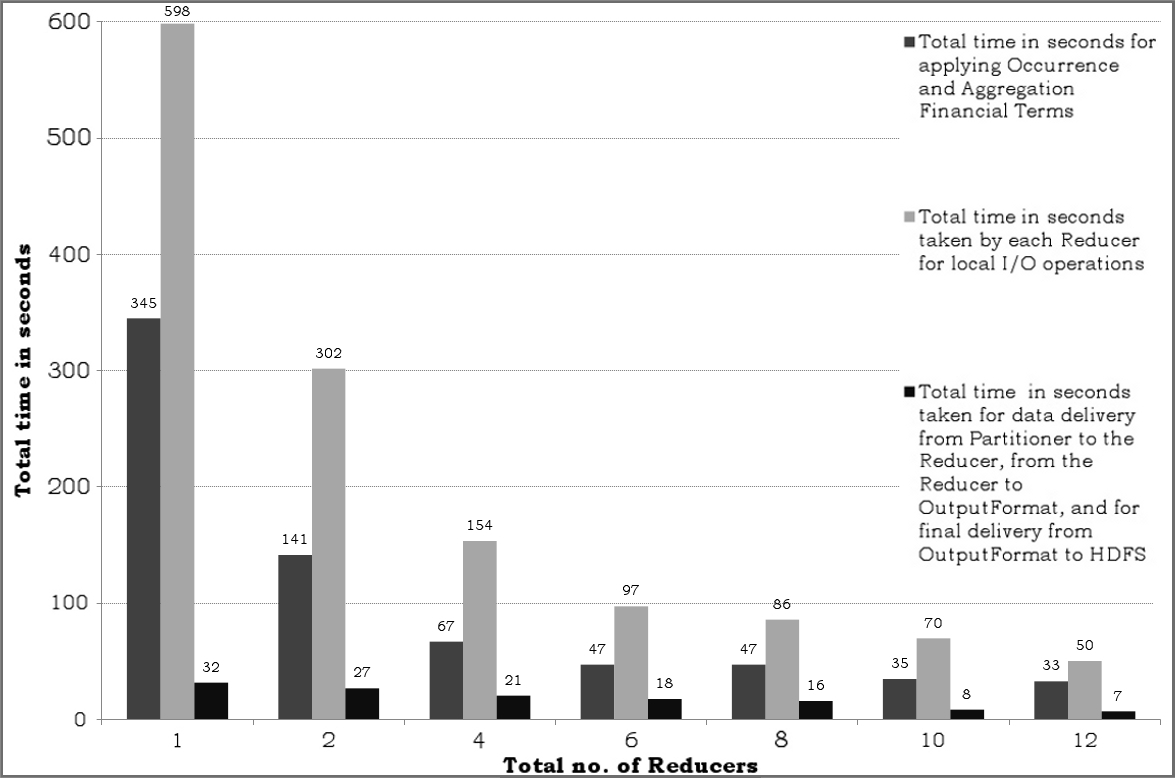}} \\
\caption{Profiled time for the first MapReduce round in the Hadoop implementation}
\label{figure3}
\end{figure}

Figure \ref{figure3} shows two bar graphs in which the time for the first MapReduce round is profiled. For the Mapper the time taken into account is for (a) applying Financial Terms, (b) local I/O operations, and (c) data delivery from the HDFS to the InputFormat, from the InputFormat to the Mapper, and from the Mapper to the Partitioner. For the Reducer the time taken into account is for (a) applying Occurrence and Aggregate Financial Terms, (b) local I/O operations, and (c) data delivery from the Partitioner to the Reducer, from the Reducer to the OutputFormat and from the OuputFormat to HDFS. For both the Mappers and the Reducers it is observed that over half the total time is taken for local I/O operations. In the case of the Mapper the mathematical computations only take $1/4^{th}$ the total time, and the total time taken for data delivery from the HDFS to the InputFormat, and from the InputFormat to the Mapper and from the Mapper to the Partitioner is only $1/4^{th}$ the total time. In the case of the Reducer the mathematical computations take $1/3^{rd}$ the total time, whereas the total time taken for data delivery from the Partitioner to the Reducer, from the Reducer to the OutputFormat, and from the OutputFormat to HDFS is nearly $1/6^{th}$ the total time. This indicates that the local I/O operations on the cluster employed is expensive though the performance of Hadoop is exploited for both the numerical computations and for large data processing and delivery.

\begin{figure}
	\centering
	\includegraphics[width = 0.495\textwidth]{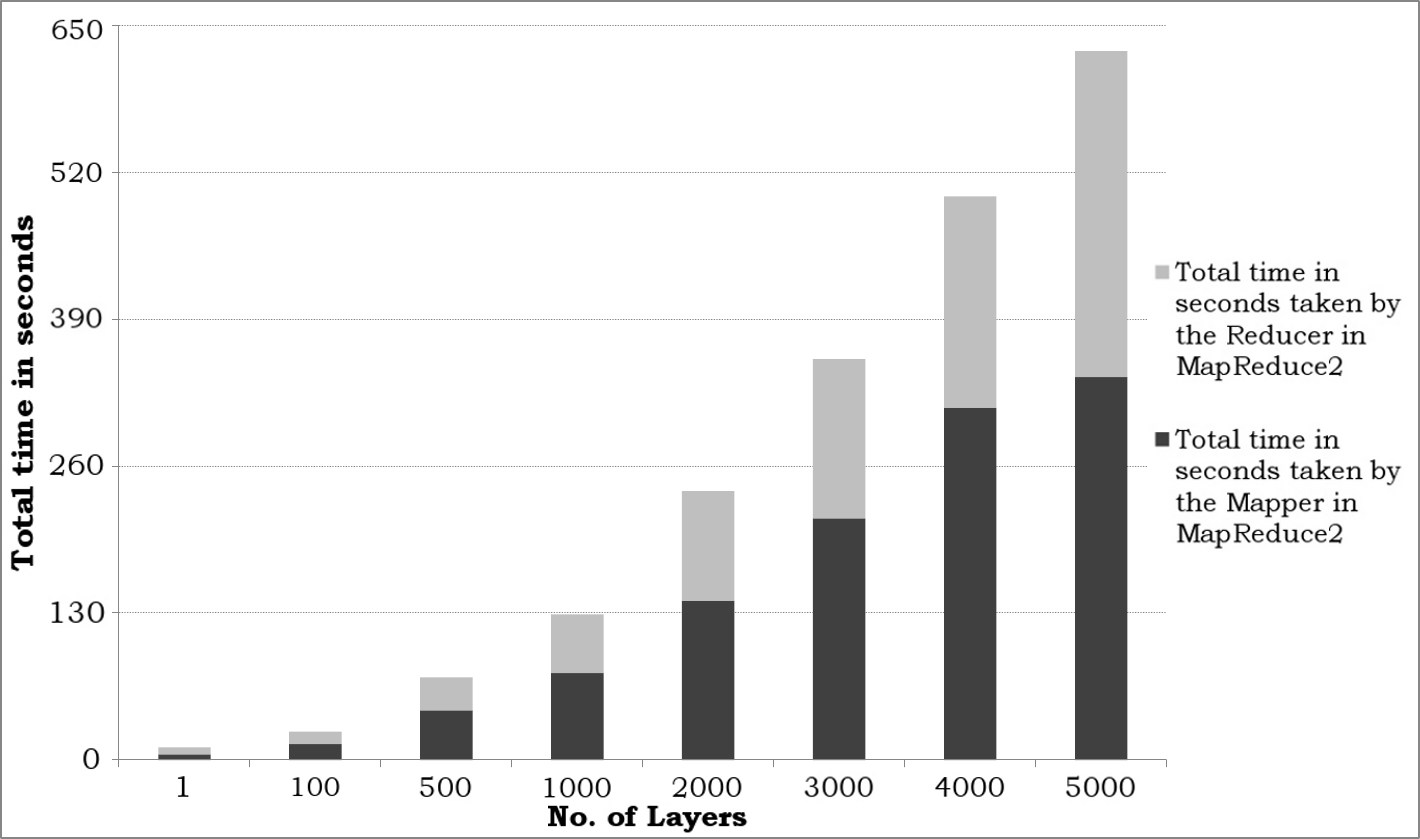}
	\caption{Number of Layers vs the total time in seconds for the second MapReduce round}
	\label{figure4}
\end{figure}

Figure \ref{figure4} shows a bar graph for the time taken in seconds for the second MapReduce round on 12 workers when the number of Layers are varied from 1 to 5000. There is a steady increase in the time taken for data processing and data delivery by the Mapper and the Reducer. Gradually the time step decreases resulting in the flattening of the trend curve. 

\begin{figure}
	\centering
	\includegraphics[width=0.495\textwidth]{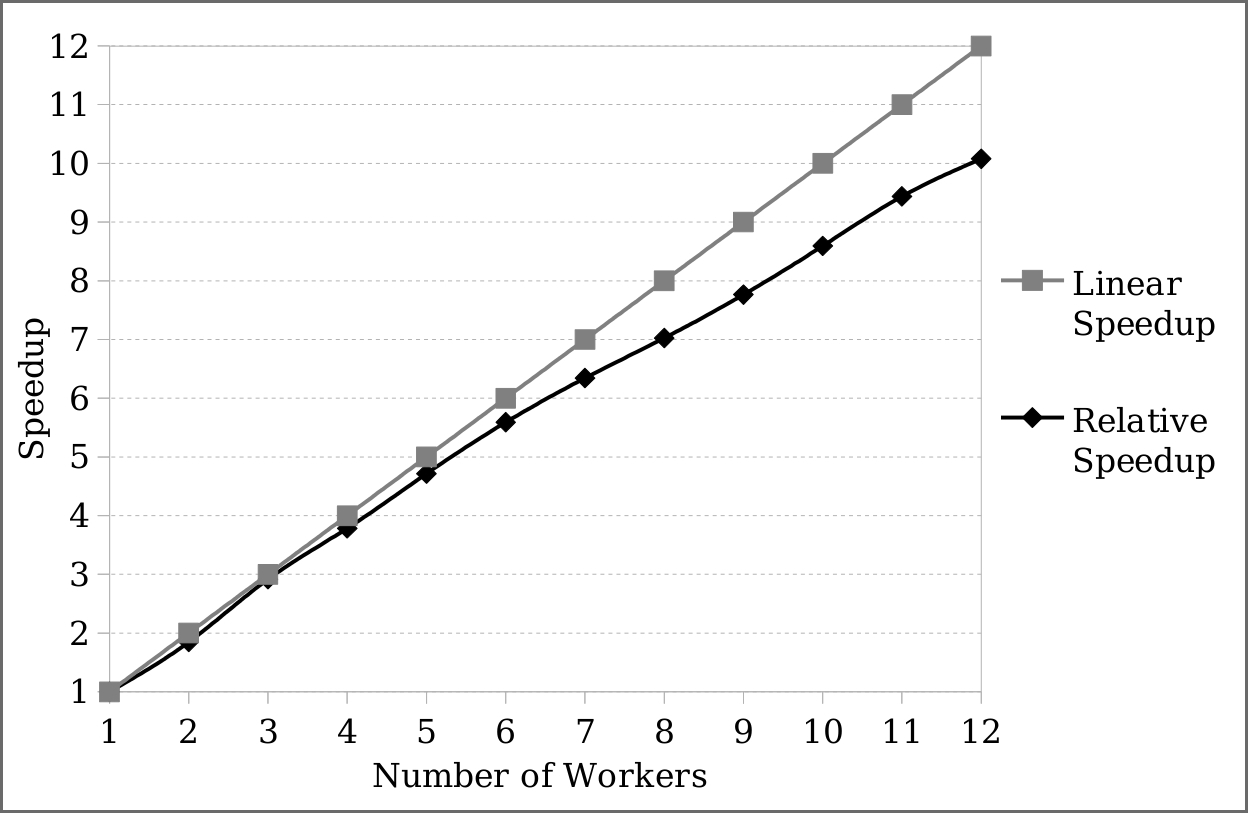}
	\caption{Speedup achieved for Aggregate Risk Analysis using MapReduce on Apache Hadoop}
	\label{figure5}
\end{figure}

\begin{figure}
	\centering
	\includegraphics[width=0.495\textwidth]{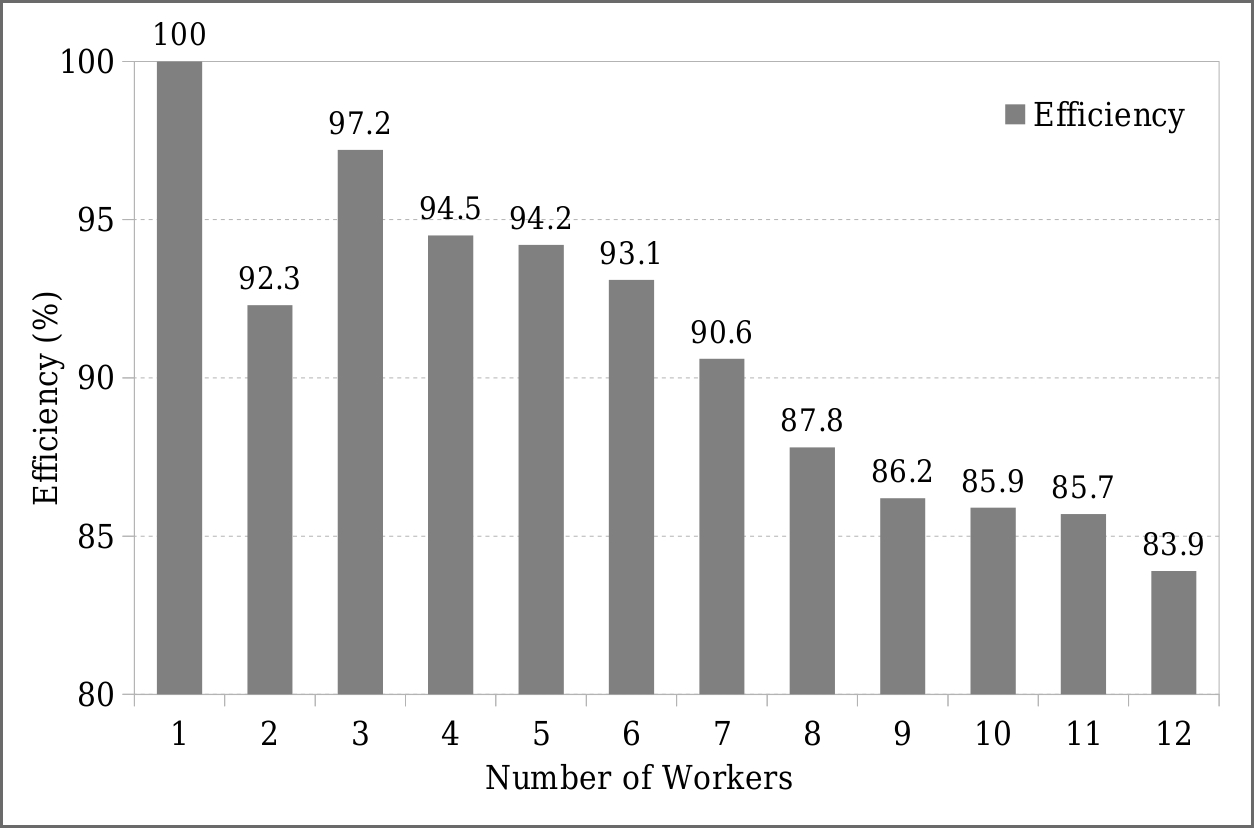}
	\caption{Efficiency achieved for Aggregate Risk Analysis using MapReduce on Apache Hadoop}
	\label{figure6}
\end{figure}

Figure \ref{figure5} shows the relative speed up achieved using MapReduce for aggregate risk analysis. There is close to linear speed up achieved until seven worker nodes are employed, and beyond seven nodes the gap between linear and relative speed up increases. This is reflected in the efficiency of the simulation for different number of workers shown in Figure \ref{figure6}. Over 90\% efficiency is achieved up to seven worker nodes. Beyond seven workers efficiency drops. For all the workers over 50\% of the time is required for local I/O operations, and around 22\% of the time is required for applying financial terms. Between 15\%-22\% of the time is required for data delivery, with a slight increase in time for each additional worker employed. This is possibly due to the overhead involved in using a centralised RAID data storage, which can be minimised if distributed file replication techniques are employed.  

The results indicate that there is scope for achieving high efficiency and speedup for numerical computations and large data processing and delivery within the Hadoop system. However, it is apparent that large overheads for local I/O operations on the workers and for data transfer onto a centralised RAID system are restraining performance. This large overhead is a resultant of the bottleneck in the connectivity between the worker nodes and the latency in processing data from local drives and the redundant data transfer to the centralised RAID storage. Therefore, efforts need to be made towards reducing the I/O overhead and seeking alternative distributed strategies to incorporate data replication for exploiting the full benefit of the Hadoop MapReduce model. 

\pagebreak
\section{Conclusion}
\label{conclusion}

Simulations for the analysis of portfolios of catastrophic risk need to manage and process large volumes of data in the form of a Year Event Table and Event Loss Tables. To be able to employ the simulations in real-time the algorithms need to rapidly process the data which poses both computational and data management challenges. In this paper, how the MapReduce model using the Hadoop framework can meet the requirements of rapidly consuming large volumes of data for the analysis of portfolios of catastrophic risk to address the challenges has been presented. An embarrassingly parallel algorithm for aggregate risk analysis is proposed and implemented using the Map and Reduce functions on the Apache Hadoop framework. The data challenges can be surmounted by employing the Hadoop Distributed File System and the Distributed Cache both offered by Apache Hadoop. A simulation of aggregate risk employing 100,000 trials with 1000 catastrophic events per trial performed on multiple worker nodes using two MapReduce rounds takes less than 6 minutes. The experimental results show the feasibility of employing MapReduce for parallel numerical computations and data management of aggregate risk analysis in real-time. 

Future work will be directed towards optimising the implementation for reducing the local I/O overheads to achieve better speedup. Efforts will be made towards incorporating additional financial filters, such as secondary uncertainty for fine-grain analysis of aggregate risk. 

\bibliographystyle{abbrv}

\begin{thebibliography}{00}

\bibitem{1}
G. Connor, L. R. Goldberg and R. A. Korajczyk, ``Portfolio Risk Analysis,'' Princeton University Press, 2010.

\bibitem{2}
A. Melnikov, ``Risk Analysis in Finance and Insurance,'' Second Edition, CRC Press, 2011.

\bibitem{2a}
A. K. Bahl, O. Baltzer, A. Rau-Chaplin and B. Varghese, ``Parallel Simulations for Analysing Portfolios of Catastrophic Event Risk,'' Workshop of the International Conference for High Performance Computing, Networking, Storage and Analysis (SC), 2012.

\bibitem{3}
J. Dean and S. Ghemawat, ``MapReduce: Simplified Data Processing on Large Clusters,'' Communications of the ACM, Vol. 51, No. 1, 2008, pp. 107-113.

\bibitem{s1}
R. R. Anderson and W. Dong, ``Pricing Catastrophe Reinsurance with Reinstatement Provisions Using a Catastrophe Model,''  Casualty Actuarial Society Forum, Summer 1998, pp. 303-322.

\bibitem{s2}
G. G. Meyers, F. L. Klinker and D. A. Lalonde, ``The Aggregation and Correlation of Reinsurance Exposure,'' Casualty Actuarial Society Forum, Spring 2003, pp. 69-152.

\bibitem{s4}
W. Dong, H. Shah and F. Wong, ``A Rational Approach to Pricing of Catastrophe Insurance,'' Journal of Risk and Uncertainty, Vol. 12, 1996, pp. 201-218. 


\bibitem{4}
K. -H. Lee, Y. -J. Lee, H. Choi, Y. D. Chung and B. Moon, ``Parallel Data Processing with MapReduce: A Survey,'' SIGMOD Record, Vol. 40, No. 4, 2011, pp. 11-20.

\bibitem{5}
T. Condie, N. Conway, P. Alvaro, J. M. Hellerstein, K. Elmeleegy and R. Sears, ``MapReduce Online,'' EECS Department, University of California, Berkeley, USA, Oct 2009, Technical Report No. UCB/EECS-2009-136. 

\bibitem{6}
D. Cummins, C. Lewis and R. Phillips, ``Pricing Excess-of-Loss Reinsurance Contracts Against Catastrophic Loss,'' The Financing of Catastrophe Risk, Editors: K. A. Froot, University of Chicago Press, 1999, pp. 93-148.

\bibitem{7}
Y. -S. Lee, ``The Mathematics of Excess of Loss Coverages and Retrospective Rating - A Graphical Approach,'' Casualty Actuarial Society Forum, 1988, pp. 49-77.

\bibitem{8}
G. Woo, ``Natural Catastrophe Probable Maximum Loss,'' British Actuarial Journal, Vol. 8, 2002.

\bibitem{9}
M. E. Wilkinson, ``Estimating Probable Maximum Loss with Order Statistics,'' Casualty Actuarial Society Forum, 1982, pp. 195-209. 

\bibitem{10}
A. A. Gaivoronski and G. Pflug, ``Value-at-Risk in Portfolio Optimisation: Properties and Computational Approach,'' Journal of Risk, Vol. 7, No. 2, 2005, pp. 1-31.

\bibitem{11}
P. Glasserman, P. Heidelberger and P. Shahabuddin, ``Portfolio Value-at-Risk with Heavy-tailed Risk Factors,'' Mathematical Finance, Vol. 12, No. 3, 2002, pp. 239-269.

\bibitem{12}
T. White, ``Hadoop: The Definitive Guide,'' 1st Edition, O'Reilly Media, Inc., 2009.

\bibitem{13}
Apache Hadoop Project: http://hadoop.apache.org/ [Last Accessed: 10 April, 2013]

\bibitem{14}
K. Shvachko, K. Hairong, S. Radia and R. Chansler, ``The Hadoop Distributed File System,'' Proceedings of the 26th IEEE Symposium on Mass Storage Systems and Technologies, 2010, pp. 1-10.

\bibitem{15}
Amazon Elastic MapReduce (EMR): http://aws.\\amazon.com/elasticmapreduce/ [Last accessed: 10 April, 2013]

\bibitem{16}
Google MapReduce: https://developers.google.com/\\appengine/docs/python/dataprocessing/overview [Last accessed: 10 April, 2013]

\end{thebibliography}

\end{document}